\documentclass[epj,final]{svjour}
\usepackage{graphics}
\usepackage{amssymb}
\usepackage{epsfig}
\newcommand{\di}{\displaystyle}
\newcommand{\mN}{M_{\scriptstyle N}}
\newcommand{\scN}{{\scriptscriptstyle N}}

\newcommand{\GeV}{\; \mathrm{GeV}}
\newcommand{\MeV}{\; \mathrm{MeV}}

\begin{document}

\authorrunning{G. Vereshkov, O.Lalakulich}
\titlerunning{Soft photons in $eN$ scattering}

\title{Logarithmic corrections and soft photon phenomenology in the multipole
model of the nucleon form factors}
\subtitle{}
\author{G. Vereshkov 
\and O.Lalakulich\thanks{\emph{Present address:} Department of Subatomic and Radiation Physics, Ghent Unibversity, Ghent}%
}                     
%
%
\institute{Research Institute of Physics,
Southern Federal University, 344090 Rostov-on-Don, Russia}
\date{}
%
\abstract{
We analyzed the presently available experimental data on  nucleon
electromagnetic form factors within a multipole model based on
dispersion relations. A good fit of the data is achieved by
considering the coefficients of the multipole expansions as
logarithmic functions of the momentum transfer squared. The
superconvergence relations, applied to this coefficients, makes the
model agree with unitary constraints and pQCD asymptotics for the
Dirac and Pauli form factors. The soft photon emission is proposed
as a mechanism responsible for the difference between the
Rosenbluth, polarization and beam--target--asymmetry data. It is
shown, that the experimentally measured cross sections depend not
only on the Dirac and Pauli form factors, but also on the average
number of the photons emitted. For proton this number is shown to be
different for different types of experimental measurements and then
estimated phenomenologically. For neutron the same mechanism
predicts, that the data form different types of experiments must
coincide with high accuracy. A joint fit of all the experimental
data reproduce the $Q^2-$dependence with the accuracy
$\chi^2/dof=0.86$. Predictions of the model, that 1) the ratios of
the proton form factors $G_E/G_M$ are different for Rosenbluth,
polarization and beam--target--asymmetry experiments and 2) similar
ratios are nearly the same for neutron, can be used for experimental
verification of the model.
} 
\maketitle

\section{Introduction}

The nucleon elastic form factors are of fundamental importance for
understanding the electromagnetic structure of the nucleon. Until recently
they only have been measured through Rosenbluth technique in experiments
on non-polarized elastic electron--nucleon scattering. This method gives
non-interfering electric $G_{Ep}$, $G_{En}$ and magnetic $G_{Mp}$,
$G_{Mn}$ form factors for proton and neutron, respectively.

Recent progress in experimental technique made it possible to use
polarized beams and/or polarized targets, which brings two new methods
\cite{Day:2007ed}, polarization transfer and beam--target asymmetry, to
determine the ratio of the electric to magnetic form factors.

New experimental data, recently obtained by these methods, show an
excellent agreement with the old Rosenbluth data at low momentum transfer,
$Q^2 \lesssim 1 \GeV^2$, and posed intriguing questions at higher $Q^2$.

\begin{figure}[htb]
        \epsfig{figure=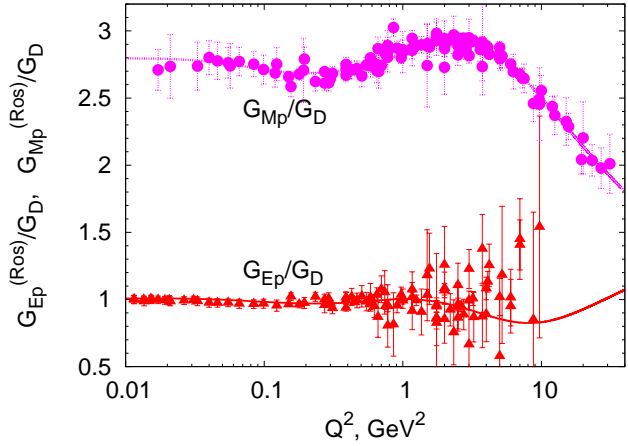,angle=-90,width=\linewidth}
        \caption{Proton form factors, obtained by Rosenbluth separation technique,
                  normalized on the dipole function $G_D=(1+Q^2/0.71)^{-2}$}
        \label{fig:Gprot}
\end{figure}

\begin{figure}[htb]
        \epsfig{figure=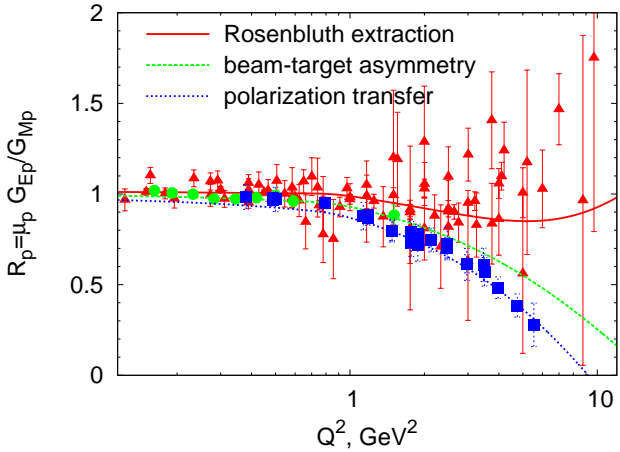,angle=-90,width=\linewidth}
        \caption{Ratio of the proton form factors, obtained with different experimental  techniques}
        \label{fig:Rprot}
\end{figure}

\begin{figure}[htb]
        \epsfig{figure=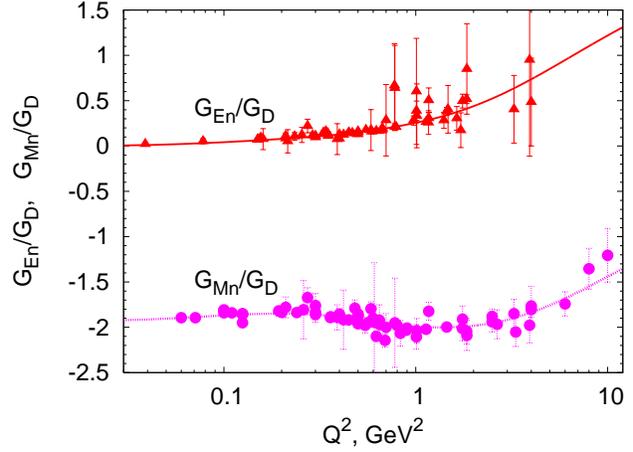,angle=-90,width=\linewidth}
        \caption{Neutron form factors, normalized on the dipole function $G_D=(1+Q^2/0.71)^{-2}$}
        \label{fig:Gneut}
\end{figure}

\begin{figure}[htb]
        \epsfig{figure=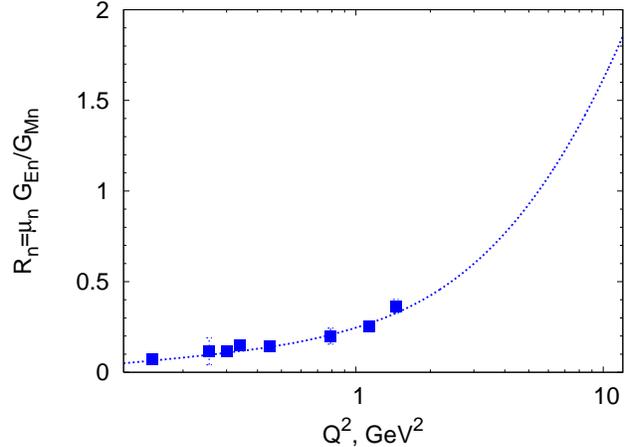,angle=-90,width=\linewidth}
        \caption{Ratio of the neutron form factors }
        \label{fig:Rneut}
\end{figure}

Proton electric and magnetic form factors, normalized on the dipole
function $G_D=(1-t/0.71)^{-2}$, as they are determined with the
Rosenbluth technique  are shown in Fig.~\ref{fig:Gprot}. The ratios
of proton form factors, as they are determined from different
experimental techniques are shown in Fig.~\ref{fig:Rprot}. Neutron
electric and magnetic form factors, also normalized on the dipole
function $G_D=(1-t/0.71)^{-2}$, come from the Rosenbluth and
polarization transfer techniques and are shown in
Fig.~\ref{fig:Gneut}.  The ratio of the neutron form factors, as it
is determined from polarization data is  shown in
Fig.~\ref{fig:Rneut}. Fit of these data within the framework of the
model, proposed in this paper, is shown by solid lines.  The
references on the original experiments are summarized in
Tables~\ref{tab:FFp},\ref{tab:FFn} (see page~\pageref{tab:FFp}, also
\cite{Friedrich:2003iz,Tomasi-Gustafsson:2005kc} for earlier
compilations).


In Figs.~\ref{fig:Gprot},~\ref{fig:Rprot},~\ref{fig:Gneut} we use
logarithmic scale for $Q^2$. This way one could easily notice, that

(i) in the whole experimentally investigated $Q^2$ region the global
$Q^2-$evolution of the form factors $G^{(Ros)}_{Ep}$,
$G^{(Ros)}_{Mp}$, $G^{(Ros)}_{Mn}$, obtained via the Rosenbluth
extraction
is described by the dipole model with an accuracy of about
$10\,-\,20\,\%$;

(ii) deviations from the dipole model are of logarithmic form;

(iii) The ratio of the proton form factors
$R_p^{(pol)}=\mu_pG^{(pol)}_{Ep}/G^{(pol)}_{Mp}$, measured in the
polarization transfer experiments, steeply falls down as $Q^2$ increases
up to $5.55 \GeV^2$. The ratio
$R^{(Ros)}_p=\mu_pG^{(Ros)}_{Ep}/G^{(Ros)}_{Mp}$, calculated from
Rosenbluth data is approximately constant. This is illustrated in
Fig.\ref{fig:Rprot}, which also includes data form beam--target asymmetry
experiments. Thus, recoil polarization measurements contradict the
Rosenbluth measurements and there is a dramatic problem to make them agree
with each other \cite{Arrington:2002cr}.


Phenomenological fit of the form factors without discussing their physical
nature is available in Refs. \cite{Bodek:2003ed,Budd:2003wb,Kelly:2004hm}.
Theoretically the form factors are investigated within different models
(their classification, short review and references see in
\cite{Punjabi:2005wq}[Section V]).

1. In low $Q^2$ region a good  description of the data is provided in the
constituent quark models, including cloudy bag and diquark models. These
models rely on the spectroscopic data to fix their free parameters and
then provide predictions for the nucleon form factors (see
\cite{Punjabi:2005wq} for a short review).

2. In high $Q^2$ region ($Q^2>20-30\GeV^2$ for form factor problem)
perturbative QCD gains its power, which makes it possible to
calculate the pQCD asymptotics for $F_{1n,p}$
\cite{Brodsky:1974vy,Lepage:1980fj} and $F_{2p,n}$
\cite{Belitsky:2002kj,Brodsky:2003pw} from pure theoretical
considerations.

3. An attempt to extend these calculations to a lower-$Q^2$ region was
recently made by Guidal {\it et. al} \cite{Guidal:2004nd} in a model,
based on nonperturbative generalised parton distributions. This model
contains free parameters, which are adjusted to fit the data on the form
factors.

4. The models at the nucleon--meson level by construction intend to
describe the data in both low- and high-$Q^2$ regions. Complying
with the low-$Q^2$ and pQCD asymptotics of the form factors must be
inevitable feature of such models. In vector meson dominance models
\cite{Gari:1992tq,Lomon:2001ga,Lomon:2002jx,Iachello:2004ki,Bijker:2004yu} the
photon couples to the nucleon via vector mesons; to date four
different vector mesons are used in such models;

5. In a model of Mainz--Bonn--Julich group
\cite{Mergell:1995bf,Hammer:1996kx,Hammer:2003ai}, based on the
dispersion--theoretical analysis, some poles of the multipole expansions
are also related to the vector meson masses. In addition the effect of
many--meson exchange is taken into account. In this approach at least
three isovector and three isoscalar mesons must be considered in order to fit
all the data.

Different approaches to the physical nature of the form factors are
complementary, supporting the idea of quark--hadron duality. It is known
nowadays, that the effective theory of hadronic fields can be derived from
QCD. This advances the idea of a univocal correspondence between the quark
and nucleon--meson models of the form factors. At present, however, such
correspondence can hardly be traced, because the calculations in
nonperturbative QCD are too complicated as well as the final form of the
nucleon--meson phenomenology is not yet established.

In this situation we base our model on two the most reliable general
consequences of the field theory. They are: (i) the multipole structure of
the form factors, which is predicted by the dispersion relations
\cite{Clementel:1956,Bergia:1961}; (ii) the asymptotical behavior of the
Dirac and Pauli form factors, calculated in perturbative QCD
\cite{Brodsky:1974vy,Lepage:1980fj,Belitsky:2002kj,Brodsky:2003pw}.

In Section~\ref{model} we introduce our formalism and establish the
full set of superconvergence relations imposed on the coefficients
of the multipole expansions in order to adjust the multipole form
factors with the QCD asymptotics and unitarity constraints. As we
have already mentioned, a similar approach was first used in the
dispersion--theoretical analysis of the nucleon form factors
developed by Mainz--Bonn--Julich group
\cite{Mergell:1995bf,Hammer:1996kx,Hammer:2003ai} since 1996. We
keep the same lines, but use improved logarithmic dependences of the
multipole coefficients. Instead of one universal logarithmic
function for the pole contributions and two more for the two-pion
continuum in the Mainz--Bonn--Julich model, we introduce four
functions with leading and subleading logarithms, which are
motivated by the modern pQCD asymptotics. This approach allows us to
make all the poles correspond to the masses of the physical mesons
in the PDG tables, as well as to achieve a better fit of the data.

Another purpose of the paper is to resolve the  above described
discrepancy in the proton form factors and propose a way to reconcile the
three sets of data. Up to now this question remained open in all the above
cited models, where fits have to be based on the recoil polarization data,
practically ignoring the high-$Q^2$ Rosenbluth data.  Much attention in
this respect was devoted to the radiative corrections
\cite{Maximon:2000hm} and in particular to the two--photon exchange
\cite{Blunden:2003sp,Blunden:2005ew,Arrington:2004ae} mechanisms. An approach 
taking into account the hadronic contribution to the electron structure functions 
is presented in \cite{Bystritskiy:2007hw}.  A global analysis of the Rosenbluth 
data \cite{Tvaskis:2005ex} found, however,
hardly ever evidence for the two--photon exchange effects. We propose,
that the physical phenomena responsible for the discrepancy is the soft
photon emission, which proceeds differently in different types of
experiments.

In Section~\ref{softphotons} we argue, that in electron--nucleon
scattering, the soft photons that are not detected experimentally
can be emitted, and this effect influence the values of the
experimentally measured electric and magnetic form factors and their
ratios. It is shown that the observables, extracted from the cross
sections, depend not only on the Dirac and Pauli form factors, but
also on the average number of the photons emitted. This dependence
results from the fact \cite{Gould:1979pw}, that the Dirac and Pauli
form factors contribute to the emission of photons differently. In
the polarization transfer experiments only events without photon
emission are experimentally possible, because otherwise the photon
would carry away the angular momentum and thus prevent the transfer
of polarization. Non--polarized electrons in the final states, on
the other hand, are inevitably accompanied by soft photon emission.
We provide formulas for the observable values in all the three types
of experiments.

Section~\ref{fit} is devoted to the numerical analysis. The joint
fits of all experimental data available is performed according to
the formulas presented in the previous sections.  The agreement
between the theory and the experimental data is achieved at the
level $\chi^2/dof=0.86$. In Section~\ref{predictions} we give our
predictions, which can be used for the experimental verification of
the model and in Section~\ref{conclusions} we summarize our
findings.

\section{Multipole model for Dirac and Pauli form factors. \label{model}}

\subsection{Formulation of the model}

The main contribution to the dispersion--relation form factors is given by
multipole expansion, which accounts for one--meson exchange in the
approximation of the narrow meson width. Retaining the isotopic symmetry,
the most general structure for the form factors at $t=-Q^2<0$ is:

\begin{equation}
\begin{array}{c}
\displaystyle
F_{1p,n}(t)=Q_{p,n}(t)-\frac12\left[\sum_{k=1}^{N_s}a_{s(k)}(t)\pm
\sum_{k=1}^{N_v}a_{v(k)}(t)\right]+
\\[4mm]
\displaystyle
+\frac12\left[\sum_{k=1}^{N_s}\frac{a_{s(k)}(t)m^2_{s(k)}}{m^2_{s(k)}-t}\pm
\sum_{k=1}^{N_v}\frac{a_{v(k)}(t)m^2_{k(v)}}{m^2_{v(k)}-t}\right],
\\[6mm]
\displaystyle
F_{2p,n}(t)=\mu^{(an)}_{p,n}(t)-\frac12\left[\sum_{i=1}^{N_s}b_{s(k)}(t)\pm
\sum_{k=1}^{N_v}b_{v(k)}(t)\right]+
\\[4mm]
\displaystyle
+\frac12\left[\sum_{k=1}^{N_s}\frac{b_{s(k)}(t)m^2_{s(k)}}{m^2_{s(k)}-t}\pm
\sum_{k=1}^{N_v}\frac{b_{v(k)}(t)m^2_{v(k)}}{m^2_{v(k)}-t}\right].
\end{array}
\label{4}
\end{equation}
Here $N_v$ and $N_s$ are, correspondingly,  the numbers of isotriplet and
isosinglet poles  of the scattering amplitude. The upper and the lower
signs correspond to proton and neutron, respectively.

Generally the phenomenology of the dispersion relations allows for the
coefficients of the multipole expansions
$Q_{p,n}(t),\,\mu^{(an)}_{p,n}(t)$, $a_{s,v(k)}(t)$, $b_{s,v(k)}(t)$ to be
slow variable (logarithmic) functions of $t$. At $-t\to 0$ the limiting
values of functions $Q_{p,n}(t)$, $\mu^{(an)}_{p,n}(t)$ are equal to the
charges and anomalous magnetic moments of nucleons:
\begin{equation}
\begin{array}{c}
\displaystyle Q_p(0)=1,\qquad \mu^{(an)}_p(0)\equiv \mu_p-1=1.793,
\\[3mm]
\displaystyle
  Q_n(0)=0,\qquad \mu^{(an)}_n(0)\equiv \mu_n=-1.913.
\end{array}
\label{Q2zero}
\end{equation}
The asymptotical behavior of the Dirac and Pauli form factors at $-t\to
\infty$ traces back to the results of the perturbative QCD:
\begin{equation}
\begin{array}{c}
\displaystyle
 F_{1p,n}(t)\to
\left(\frac{4\mN^2}{t}\right)^2\frac{C_{1p,n}}{\ln^{p_1}|t|/\Lambda^2},
\\ \di
F_{2p,n}(t) \to
\left(\frac{4\mN^2}{t}\right)^3 \frac{C_{2p,n}}{\ln^{p_2}|t|/\Lambda^2},
\\[3mm]
\displaystyle p_1=2+\frac{32}{9\beta},\quad
p_2=\frac{8}{3\beta},\quad \beta=11-\frac23n_f.
\end{array}
\label{pQCD}
\end{equation}
where $\Lambda\simeq 300 \MeV$ is the QCD scale. The exponent $p_1$ is
calculated theoretically and is know with rather good accuracy; the
constants $C_{1p,n}$ are to be evaluated by comparing the experimental
data with the parton model
\cite{Brodsky:1974vy,Lepage:1980fj,Brodsky:2003pw}. The asymptotic of the
Pauli form factor from Eq. (\ref{pQCD}) and the constants $C_{2p,n}$ are
evaluated in \cite{Belitsky:2002kj}. Another asymptotic is proposed in
\cite{Brodsky:2003pw}.

The power dependence of the asymptotics in Eq.~(\ref{pQCD}) results from
the quark counting rules \cite{Matveev:1973ra,Brodsky:1974vy}. Those in
turn follow from the analysis of the process with the quark momentum
redistribution among the quark and gluon components of the nucleon. The
logarithmic functions describe the renormalization of the color charge and
the wave functions of the partons. Note, that in the elastic scattering
theory the region of the perturbative QCD applicability is estimated
differently than in the deep inelastic scattering. We discuss this
question in the next paragraph before proceeding with the form factors.

In DIS the perturbative description of the inclusive processes is
possible, if for the initial act of the electron--parton interaction  the
inequality $|t|\gg T_g^{-2}$ is satisfied, with $T_g\simeq (1.5\GeV)^{-1}$
being the correlation length (the characteristic scale) of the
nonperturbative gluon fluctuations. In case of the elastic $eN-$scattering
the momentum, transferred to one proton in the initial interaction act, is
uniformly distributed among all partons; and in the region of QCD
applicability all processes of momentum redistribution must be hard. The
number of partons, participating in these processes, is no less than the
number of the valent quarks $n_q=3$, so the perturbative description of
the elastic scattering is only justified for $|t|\gg
|t_{\scriptscriptstyle QCD}|$, with $|t_{\scriptscriptstyle QCD}|\simeq
n_q T_g^{-2}\simeq 7 \GeV^{2}$. Among the experimental data available,
only the proton magnetic form factor at $|t|= 20\,-\,30\, \GeV^2$ could
probably be compared with the perturbative QCD.  However, the very fact
that the QCD asymptotics (\ref{pQCD}) exist is of fundamental importance
at discussing the general mathematical structure of the form factors.

\subsection{The Superconvergence Relations \label{sumrules} }

Thus, multipole form factors (\ref{4}) must satisfy the asymptotics
(\ref{pQCD}), which automatically include the unitarity conditions. In
this way one phenomenologically accounts for the QCD effects at the valence
quark level, that is (i) the redistribution of the transferred momentum
between quarks, (ii) subsiding of chromodynamical interactions as $Q^2$
increases, (iii) dependence of parton (valence quark) distribution
functions on $Q^2$.

There are two ways to satisfy pQCD asymptotics. One can nullify the
non--pole terms in the expansion (nullify the so called "nucleon core")
and impose the corresponding conditions to each of the coefficients
$a_{s,v(k)}(t)$, $b_{s,v(k)}(t)$ by modelling them by pow--low and
logarithmic functions. This way is used in
Refs.~\cite{Lomon:2001ga,Lomon:2002jx}.

Another way is to satisfy the pow--low asymptotics by imposing the
"superconvergence relations", as they are referred to in the papers of the
Mainz--Bonn-Julich  group, over the whole sets of the multipole
coefficients, so that they relate parameters of the different poles to
each other. The multipole coefficients in these papers are considered as
logarithmic functions with the leading logarithms only.

The same approach was taken later by Bratislava group
in~\cite{Adamuscin:2002ca}, who use the term "asymptotic conditions". The
logarithmic dependences of the multipole coefficients are, however,
ignored in the papers of this group, which makes a good fit of the
experimental data difficult.

We implement the latter way and impose the superconvergence relations(SR)
on the four sets of our coefficients $a_{s,v(k)}(t)$, $b_{s,v(k)}(t)$. Let
us expand Eq.(\ref{4}) on powers $1/t$ and nullify the terms of the order
$t^0$, $t^{-1}$ for $F_{1p,n}$ and $t^0$, $t^{-1}$, $t^{-2}$ for
$F_{2p,n}$. Since isovector and isoscalar modes are independent, there are
10 independent SR:
\begin{equation}
\begin{array}{c}
\displaystyle \sum_{k=1}^{N_{s,v}}a_{s,v(k)}(t)=q_{s,v}(t), \qquad
\sum_{k=1}^{N_{s,v}}a_{s,v(k)}(t)m^2_{s,v(k)}=0,
\\[3mm]
\displaystyle \sum_{k=1}^{N_{s,v}}b_{s,v(k)}(t)=(\mu_p\pm\mu_n-1)\mu_{s,v}(t),
\\[3mm]
\displaystyle \sum_{k=1}^{N_{s,v}}b_{s,v(k)}(t)m^2_{s,v(k)}=0,\qquad
\sum_{k=1}^{N_{s,v}}b_{s,v(k)}(t)m^4_{s,v(k)}=0,
\end{array}
\label{7}
\end{equation}
where
\begin{equation}
\di q_{s,v}(t)\equiv Q_p(t)\pm Q_n(t), \quad
\mu_{s,v}(t)\equiv\frac{\mu^{(an)}_p(t)\pm
\mu^{(an)}_n(t)}{\mu_p\pm\mu_n-1}.
\label{8}
\end{equation}
The upper and lower signs here correspond to the isoscalar ($s$) and
isovector ($v$) parts, respectively. The SR (\ref{7}) and the
normalization conditions (\ref{Q2zero}) allow us to attribute the
$t$ dependence of the coefficients $a_{s,v(k)}(t)$, $b_{s,v(k)}(t)$
to the four phenomenological logarithmic functions $q_{s,v}(t)$,
$\mu_{s,v}(t)$ (which do not depend on $k$) and introduce new
parameters $\tilde a_{s,v(k)}$, $\tilde b_{s,v(k)}$:
\[
\begin{array}{c}
\displaystyle a_{s,v(k)}(t)= q_{s,v}(t)\tilde a_{s,v(k)},
\\[2mm]
\di b_{s,v(k)}(t)=(\mu_p\pm\mu_n-1)\mu_{s,v}(t)\tilde b_{s,v(k)}.
\end{array}
\]
In terms of these parameters the SR are reduced to the following
expressions:
\begin{equation}
\begin{array}{c}
\displaystyle \sum_{k=1}^{N_{s,v}}\tilde a_{s,v(k)}=1,\qquad
\sum_{k=1}^{N_{s,v}}\tilde a_{s,v(k)}m^2_{s,v(k)}=0,
\\[4mm]
\displaystyle \sum_{k=1}^{N_{s,v}}\tilde b_{s,v(k)}=1,\qquad
\sum_{k=1}^{N_{s,v}}\tilde b_{s,v(k)}m^2_{s,v(k)}=0,
\\[4mm] \di
\sum_{k=1}^{N_{s,v}}\tilde b_{s,v(k)}m^4_{s,v(k)}=0.
\end{array}
\label{SR}
\end{equation}
The form of the functions $q_{s,v}(t)$ and $\mu_{s,v}(t)$ will be
discussed in the next subsection and finally given in
Eq.~(\ref{qmu}).

 Notice, that SR
(\ref{SR}) are independent of the kinematic variable $t$, so the values
$\tilde a_{s,v(k)}$, $\tilde b_{s,v(k)}$ can be considered as numbers. It
is also important, that three SR for the parameters $\tilde b_{v(k)}$ are
only compatible with three or more isovector and isoscalar mesons. Applying SR
(\ref{SR}) to the form factors (\ref{4}) makes the non--pole terms (the
first and the second terms) vanish and results in:

\begin{equation}
\begin{array}{c}
\displaystyle F_{1p,n}(t)=
\frac12\left[q_{s}(t)\sum_{k=1}^{N_s}\frac{\tilde a_{s(k)}m^2_{s(k)}}{m^2_{s(k)}-t}
 \pm q_{v}(t)\sum_{k=1}^{N_v}\frac{\tilde
a_{v(k)}m^2_{v(k)}}{m^2_{v(k)}-t}\right],
\\[5mm]
\displaystyle F_{2p,n}(t)=
\frac12\left[(\mu_p+\mu_n-1)\mu_{s}(t)\sum_{k=1}^{N_s}
\frac{\tilde b_{s(k)}m^2_{s(k)}}{m^2_{s(k)}-t}
\pm\right.
\\[3mm]
\displaystyle \left.\pm
(\mu_p-\mu_n-1)\mu_{v}(t)\sum_{k=1}^{N_v}\frac{\tilde b_{v(k)}m^2_{v(k)}}{m^2_{v(k)}-t}\right].
\end{array}
\label{10}
\end{equation}

\subsection{Logarithmic corrections \label{logarithcorr}}

For the logarithmic functions (\ref{8}) their limit values at $-t\to 0$
can be derived from (\ref{Q2zero}). Their pQCD asymptotics at $-t\to
\infty$ follow from (\ref{pQCD}):
\begin{equation}
\begin{array}{c}
\displaystyle q_{s,v}(t)\to \left\{{1,\qquad\qquad\qquad -t\to 0,} \atop
\displaystyle{\frac{C_{1s,v}}{\ln^{p_1}|t|/\Lambda^2},\qquad
-t\to \infty},\right.
\\[3mm] \displaystyle
\mu_{s,v}(t)\to \left\{{1,\qquad\qquad\qquad -t\to 0,} \atop
\displaystyle{\frac{C_{2s,v}}{\ln^{p_2}|t|/\Lambda^2},\qquad
-t\to \infty},\right.
\end{array}
\label{11}
\end{equation}
where  $C_{js,v}$ will be derived below and given in Eq.(\ref{C12sv})

Making an asymptotic expansion of (\ref{10}) and using (\ref{11}) one gets
at $-t\to\infty$ the following asymptotics of the form factors:
\begin{eqnarray}
\di F_{1p,n}(t)&\to& -\frac{1}{2t^{2}\ln^{p_1}|t|/\Lambda^2}
\left[C_{1s} \sum_{k=1}^{N_s}\tilde a_{s(k)}m^4_{s(k)}
\right. \nonumber \\  & &  \left.
\pm
C_{1v}\sum_{k=1}^{N_v}\tilde a_{v(k)}m^4_{v(k)}\right]
\nonumber \\  & = & 
- \frac{ 16\mN^4}{2t^{2}\ln^{p_1}|t|/\Lambda^2}
\left[C_{1s} \kappa_{1s} \pm C_{1v}\kappa_{1v}\right], \nonumber
\\[3mm]
\di F_{2p,n}(t) &\to&
\frac{ - 1}{2t^{3}\ln^{p_2}|t|/\Lambda^2}\left[(\mu_p+\mu_n-1)C_{2s}
\sum_{k=1}^{N_s}\tilde b_{s(k)}m^6_{s(k)}
\right. \nonumber \\  &  &  \left.
 \pm (\mu_p-\mu_n-1)C_{2v}\sum_{k=1}^{N_v}\tilde
b_{v(k)}m^6_{v(k)}\right]=  \nonumber
\\[2mm] \di
&=& -\frac{64\mN^6}{2t^{3}\ln^{p_2}|t|/\Lambda^2}
\left[(\mu_p+\mu_n-1)C_{2s} \kappa_{2s} 
\right. \nonumber \\  & &  \left.
 \pm (\mu_p-\mu_n-1)C_{2v}\kappa_{2v}\right]\, .
\label{12}
\end{eqnarray}
Here we introduced four dimensionless parameters $\kappa_{j s,v}$
($j=1,2$)
\begin{equation}
\begin{array}{c} \di
\kappa_{1s,v}=\frac{1}{16\mN^4}\sum\limits_1^{N_{s,v}}\tilde a_{s,v(k)}
m_{s,v(k)}^4,
\\[2mm] \di
\kappa_{2s,v}=\frac{1}{64\mN^6}\sum\limits_1^{N_{s,v}}\tilde b_{s,v(k)}
m_{s,v(k)}^6.
\end{array}
\label{kappa}
\end{equation}
Matching (\ref{12}) with (\ref{pQCD}) yields the constants $C_{j s ,v}$
\begin{equation}
\displaystyle C_{1s,v}=-\frac{C_{1p}\pm C_{1n}}{\kappa_{1s,v}},
\quad
C_{2s,v}=-\frac{C_{2p}\pm C_{2n}}{\kappa_{2s,v}(\mu_p \pm \mu_n -1)}
\label{C12sv}
\end{equation}

Now, that the asymptotics (\ref{11}) for the logarithmic functions are
known, one could try to interpolate them in the whole range of $t$ (A
similar approach is used, for example, in Ref.\cite{Brodsky:2003pw}.) As
it shown by our numerical experiments, a good agreement with the data can
be achieved by making use of the following interpolation formulas:
\begin{equation}
\begin{array}{c}
\displaystyle q_{s,v}(t)=\left(1+ h_{1s,v}\ln\tilde t+ g_{1s,v}\ln^2\tilde
t\right)^{-p_1/2},
\\[3mm]
\di \mu_{s,v}(t)=\left(1+ h_{2s,v}\ln\tilde t+ g_{2s,v}\ln^2\tilde
t\right)^{-p_2/2},
\end{array}
\label{qmu}
\end{equation}
where $\tilde t=1+|t|/\Lambda^2$. Parameters $h_{j s,v}$ are free fit
phenomenological parameters, which correspond to the subleading logarithms
in pQCD. Parameters
\begin{equation} \di
g_{js,v}=\left(C_{js,v}\right)^{-2/p_j}
\label{paramg}
\end{equation}
correspond to the leading pQCD logarithms and are expressed
according to Eq.~(\ref{paramg}) via the other fit parameters and
constants $C_{1p,n}$, $C_{2p,n}$, which physically describe the
partonic structure of the nucleon. These constants can, in
principle, be calculated within pQCD or determined experimentally.
If the numerical values of these constants were reliably known,
parameters $g_{js,v}$ could be calculated via other fit parameters
$\kappa_{js,v}$. To date, however, the multipliers in the partonic
distribution functions, on which these constants depend, are not yet
established, their values in different references vary within a
certain ranges. So, the values $C_{1p,n}$, $C_{2p,n}$ are not
unambiguously determined. In this situation we use $g_{js,v}$ as
free fit parameters. Obtained in this way, these constants must be
in agreement with pQCD estimations.

Thus, in the model presented here, all the parameters introduced
except $h_{js,v}$ (j=1,2) have a direct physical meaning. The four
adjusting parameters $h_{js,v}$ are necessary to tie together the
low- and high $Q^2$ regions.

\subsection{Vector Dominance model as a simple realization of the multipole expansion}

In the vector dominance model the poles of the multipole expansion
in $t-$plane are preset by the masses of vector mesons $\rho$,
$\omega$, $\phi$. The numbers of poles, $N_v$ and $N_s$, are the
number of isovector ($\rho$) and isoscalar ($\omega$, $\phi$)
mesons, respectively. In most of the earlier papers only three
ground states of vector mesons, $\rho(770)$, $\omega(782)$ and
$\phi(1020)$, are taken into account; in some models
\cite{Lomon:2001ga,Lomon:2002jx} $\rho(1450)$ state is also
included.

Experimentally five $\rho\omega\phi$--families are known. The three
lightest of them are:
\begin{equation}
\begin{array}{c}
\rho(770), \quad \omega(782), \quad \phi(1019),
\\
\rho(1450), \quad \omega(1420), \quad \phi(1680),
\\
\rho(1700), \quad \omega(1650), \quad X(1750). \label{2}
\end{array}
\end{equation}

One could argue, that the photon hadronization mainly proceeds through the
vector mesons of the first family and the hadronization amplitudes
$A(\gamma^*\to V^*_k)$ are small for the other families. This is indeed
well known and follows from the experimental data on the widthes
$\Gamma(V\to e^-e^+)\sim |A(V\to \gamma^*)|^2$. In the multipole form
factors, however, each coefficient of the multipole expansions is
proportional to the hadronization amplitude multiplied on the amplitude
$A(V_k^*N\to N)$ of the virtual vector meson absorption by the target
nucleon. We define the composite amplitude as
\begin{equation}
\di A_k(eN\to eN)\sim A(\gamma^*\to V^*_k)\times
A(V_k^*N\to N).
\label{3}
\end{equation}
There is no reason to assume the composite amplitude, $A_k(eN\to eN)$, to
be small, because the small hadronization amplitude could be multiplied on
large absorption amplitude. Thus, the composite amplitudes could be
comparable for all families. Matching the theoretical model with the
experimental data, we will be able to fit the coefficients of the
multipole expansion and thus extract information on the composite
amplitudes.

The SR (\ref{SR}) are formulated for the sum of several multipole
coefficients, that is for the whole meson spectrum. For the coefficients
$b_{v,s(k)}$ there are three independent SR, which can be satisfied with
at least three vector mesons of each type (isovector and isoscalar).
Because of these SR, the number of free fit parameters in the model
considered can be fewer, than that in the models with truncated spectrum,
despite a larger number of mesons.

Another contribution to the nucleon form factors may come from light meson
continua. This topic was recently investigated by the Mainz--Bonn-Julich
group in
Ref.\cite{Hammer:2003ai,Hammer:2003qv,Belushkin:2005ds,Belushkin:2006qa}.
Within their model with only two--pion continuum \cite{Hammer:2003qv},
they obtained a good description of the data with the exception of the
JLab $G_{Ep}/G_{Mp}$ data. This is resolved in the following paper
\cite{Belushkin:2006qa}, where $\pi\pi$, $\rho\pi$ and $K\bar{K}$
contributions are included. As it will be shown in Section~\ref{fit}, in
the simplest modification of our model  with meson poles only, we reach a
similar or even a better level of accuracy for the overall fit. Since the
continua give a main contribution at low $Q^2$, while our main purpose was
to describe all the data and to resolve the discrepancy at high $Q^2$, we
decided not to complicate our model and leave this topic beyond the scope
of our paper. The price for keeping the model simple is its
non-sensitivity to the physical effects that are important at low-Q2. In
particular, beyond the scope of the model remains discussion on the role
of sea quarks in the nucleon and on nucleon radii. Subsequent development
of the model by including the meson continua can be a subject of further
investigation.

Another non-trivial intriguing questions, widely discussed in QCD, are the
nucleon spin structure \cite{Bass:2004xa,Bass:2006cr} and nucleon 
strangness \cite{Carvalho:2005nk,vonHarrach:2005at,Thomas:2006jf}. DIS
experiments show us, that quarks carry only a small fraction of the total
angular momentum of the proton and strange sea quarks in the proton are
strongly polarized opposite to the polarization of the proton. The
challenge to understand this nontrivial structure of the proton
requires studying the contributions from sea quarks, that is from the
meson cloud. The complicated structure of the nucleon should somehow
reveal itself also in elastic scattering. It would be reasonable to pose a
problem to include these effects in the phenomenological model for elastic
scattering. For these one would need to formulate the phenomenological
model for the elastic spin structure functions, to introduce elastic
strange form factors and also to consider meson cloud contributions. Such
a model would contain a large number of free fit parameters, and thus would
require more precise experimental data to fit them.

\section{Soft photons \label{softphotons}}

Let us proceed with the puzzling discrepancy between the Rosenbluth
and polarization transfer experiments on a proton target. It was
found recently, that the two-photon effects calculated within
perturbation theory appear to be too small to explain the observed
discrepancy \cite{Rekalo:2003xa,Tvaskis:2005ex}. In this situation we propose to
search for the effects beyond the perturbation theory, that is to
consider the soft bremsstrahlung accompanying the  scattering of the
charged particles. Notice, that by ''soft bremsstrahlung'' we refer here to quasiclassical 
non--perturbative effect. It should be distinguished from the far more known 
perturbative contribution, the infrared divergency of which cancels the 
correspondent divergency of the two--photon exchange.

The multiple emission of soft photons in the electron--nucleon
scattering is a quasi--classical process with a peculiar features
due to internal nucleon structure. The incoming electron scatters in
a compound field, originated from the nucleon charge (Dirac form
factor) and its magnetic moment (Pauli form factor). It was noticed
in
 \cite{Gould:1979pw}, that the soft photons are mainly produced
 in the electric field of
the Dirac form factor, while the contribution of Pauli form factor
is negligible. This disparity of the two form factors can be also
shown by direct calculation of the spectral distribution and the
total intensity of the soft bremsstrahlung. This calculation is done
in Section \ref{DandP} relying on the calculation method from
\cite[Problem 1 to \S 98]{Berestetsky:1982aq}. This calculation
shows, that two non-interfering fluxes of photons, "electric" and
"magnetic" ones, have different spectral functions and different
average number of photons. On this grounds we suppose, the soft
photon emission beyond the perturbation theory can be taken into
account by multiplying the Dirac and Pauli form factors on the
corresponding "electric" and "magnetic" emission amplitudes. The
corresponding calculations are done in Section~\ref{DPamplitudes}.

In the case of Rosenbluth type experiments, where protons in the
final state are not polarized, multiple emission of soft photons is
natural and the emission amplitudes are characterised by the average
numbers of "electric" and "magnetic" photons. In the polarisation
transfer experiments, however, any emitted photon would carry away
the angular momentum and thus prevent the transfer of polarization,
which would destroy the correlation of  the electron and proton
polarization vectors. Thus we assume, that in the polarization
transfer experiments no soft photon can be emitted.

In Sections \ref{FFRos}, \ref{FFpol} and \ref{FFbta} we show that within
our model the difference with respect to the multiple emission of the soft
photons can be responsible for the observed discrepancy and ensures that
the form factors, observed in the beam--target asymmetry experiments are
different from those in both the Rosenbluth and polarization transfer
experiments.

\subsection{Dirac and Pauli subsystems in the multiple emission of soft photons
\label{DandP}}

Let us consider the differential cross section of $eN$ scattering at the
angle $\theta$ with the emission of one photon of frequency $\omega$:
\begin{equation}
\di d\sigma=\alpha_{\rm em}
\frac{d\omega}{\omega}F(\xi(\theta))d\sigma_{Ros}(\theta).
\label{si1}
\end{equation}
The amplitude $F(\xi(\theta))$ of soft photon emission
\[
\begin{array}{c}
\displaystyle F(\xi)=
\frac{2}{\pi}\left[\frac{2\xi^2+1}{\xi\sqrt{\xi^2+1}}\ln(\xi+\sqrt{\xi^2+1})-1\right],
\\ \di \xi=\frac{\varepsilon}{m}\sin\frac{\theta}{2}
\end{array}
\]
was calculated in \cite[\S 98]{Berestetsky:1982aq} within perturbation
theory and describes the emission a soft photon integrated over the whole
range of photon angles. Here $\varepsilon$, and $m$ are correspondingly
electron energy and mass.

In one--photon approximation it was shown, that the ultrarelativistic
electrons are mainly emitted within the angle range
\begin{equation}
\di \frac{m^2\omega}{\varepsilon^3}\ll \theta \lesssim
\frac{m}{\varepsilon}.
\label{theta}
\end{equation}

For a nucleon with charge $Z$ and anomalous magnetic moment $\mu_{an}$,
the elastic cross section for such small angles is
\begin{eqnarray}
\di d\sigma_{Ros}&\approx&
\frac{4\alpha^2}{\varepsilon^2\theta^4}\left(F_1^2(t)-\frac{t}{4M_\scN^2}F_2^2(t)\right)d\Omega
\nonumber
\\
&\approx&
4\alpha^2\left(\frac{Z^2}{\varepsilon^2\theta^4}+\frac{\mu_{an}^2}{4M_\scN^2\theta^2}\right).
\label{el1}
\end{eqnarray}

The spectral distribution of the soft emission can be obtained by
substituting (\ref{el1}) into (\ref{si1}) and integrating over the
electron scattering angles. Keeping in mind, that in the region of angles
(\ref{theta}) one has $\xi\ll 1$ and $F(\xi)\approx (8/3\pi)\xi^2$, one
gets:

\begin{eqnarray}
\di d\sigma_\omega&=&\frac{16\alpha^3}{3m^2}\frac{d\omega}{\omega}
\int\limits_{\theta_{min}(\omega)}^{\theta_{max}}\left(\frac{Z^2}{\theta}+
\frac{\mu_{an}^2\varepsilon^2}{4M_\scN^2}\theta\right)d\theta=
\nonumber
\\ \di
&=& d\sigma_\omega^{(Z)}+d\sigma_\omega^{(\mu_{an})}, \nonumber
\\[5mm]
\di d\sigma_\omega^{(Z)}&=&
\frac{16Z^2\alpha^3}{3m^2}\ln\frac{\varepsilon^2}{m\omega}\frac{d\omega}{\omega},
\nonumber
\\[2mm] \di
d\sigma_\omega^{(\mu_{an})}&=&\frac{2\alpha^3\mu_{an}^2}{3M_\scN^2}\frac{d\omega}{\omega}.
\label{si2}
\end{eqnarray}

According to Eq.~(\ref{si2}) soft emission on the proton consist of
two subsystems with different intensities and different spectral
distributions. The term $d\sigma_\omega^{(Z)}$ describes the
electric dipole emission, generated in the field of the Dirac form
factor; the term $d\sigma_\omega^{(\mu_{an})}$ describes the
magnetic bremsstrahlung in the field of the Pauli form factor. As we
have already mentioned, the magnetic bremsstrahlung is strongly
suppressed in comparison with the dipole emission: \begin{equation} \di
\frac{d\sigma_\omega^{(Z)}}{d\sigma_\omega^{(\mu_{an})}}=
\frac{8Z^2M_\scN^2}{\mu_{an}^2m^2}\ln\frac{\varepsilon^2}{m\omega}\gg
1. \label{disparity} \end{equation}

However, as a matter of principle, each type of emission is
experimentally verifiable. Indeed, for the target neutron $(Z=0)$
only magnetic bremsstrahlung is possible and its characteristics
coincide with those for the proton up to a multiplier. Thus, having
neutron data, one could differentiate the two subsystems of the soft
emission for the proton. Notice also, that the soft photons from
different subsystems have different characteristic frequency
spacing. This fact can be proved by analyzing the total cross
section of the bremsstrahlung $\sigma_\gamma=\int d\sigma_\omega$,
which can be obtained by integrating (\ref{si2}) over the photon
frequencies. The result of integration, as it is conventional in the
soft photon theory, depends on two parameters: the minimal
$\omega_{min}$ and maximal $\omega_{max}$ frequencies of the soft
emission:

\begin{equation}
\begin{array}{c}
\sigma_\gamma=\sigma_\gamma^{(Z)}+\sigma_\gamma^{(\mu_{an})},
\\[5mm]
\di
\sigma_\gamma^{(Z)}=\frac{16Z^2\alpha^3}{3m^2}\left(2\ln\frac{\varepsilon}{m}
\ln\frac{\omega_{max(Z)}}{\omega_{min}}
\right.
\\[3mm] \di \hspace*{20mm}
\left.
-\frac12\ln^2\frac{\omega_{max(Z)}}{m}+
\frac12\ln^2\frac{\omega_{min}}{m}\right),
\\[5mm]
\di \sigma_\gamma^{(\mu_{an})}=
\frac{2\alpha^3\mu_{an}^2}{3M_\scN^2}\ln\frac{\omega_{max(\mu_{an})}}{\omega_{min}}.
\end{array}
\label{sitot}
\end{equation}
The minimal frequency, $\omega_{min}$, is determined by the accuracy up to
which the energies of the initial and final particles are measured by a
given detector. The very notion about the soft photons is constrained by
the approximate factorization of the scattering amplitude with one photon
emission, which is valid only in the low-frequency region. Thus, the
maximal frequencies, $\omega_{min(Z, \mu_{an})}$, inevitably appear in the
soft photon phenomenology as a cut--off parameters governed by the
conditions that the photons are soft. It follows from (\ref{sitot}), that
$\sigma_\gamma^{(\mu_{an})}<<\sigma_\gamma^{(Z)}$.

Evidently the average numbers of the soft photons emitted in the
field of the Pauli and Dirac form factors must satisfy the same
inequality $\bar n^{(\mu_{an})}<<\bar n^{(Z)}$. Besides,
$\sigma_\gamma^{(\mu_{an})}$ is of logarithmic order, while
$\sigma_\gamma^{(Z)}$ is of squared logarithmic order. It is natural
to suppose, that the average number of photons have correspondingly
the same orders of magnitude $\bar n^{(\mu_{an})}\sim \ln \omega$,
$\bar n^{(Z)} \sim \ln^2 \omega$.

When the general formula for the average number of the soft photons $\bar
n$ is derived in \cite{Berestetsky:1982aq}, \S\, 120, it is shown that the
information about the field, in which the soft emission is formed, reveals
only in the cut--off parameter $\omega_{max}$. The approximate formula
accurate to the logarithm squared  is given at the end of \S\, 120 in
\cite{Berestetsky:1982aq}:
\begin{equation}
\di
\bar n \simeq \frac{\alpha}{2\pi}\left[\ln^2\frac{|t|}{m^2}+
2\ln\frac{|t|}{m^2}\left(\ln\frac{\omega_{max}^2}{\omega_{min}^2}-
\ln\frac{\varepsilon^2}{m^2}\right)\right].
\label{sph}
\end{equation}
A more accurate expression is obtained recently in \cite{Maximon:2000hm}.
Since the average number of photons emitted in the field of the Dirac form
factor is determined with double logarithmic accuracy, the estimate for
$\bar n_{(Z)}\equiv \bar n_{1p}$  can be obtained from (\ref{sph}) with
$\omega_{max(Z)}\sim \varepsilon$:

\begin{equation}
\di
\bar n_{1p} \simeq \frac{\alpha}{2\pi}\ln^2\frac{|t|}{\omega_{min}^2}.
\label{sph1p}
\end{equation}

Further for numerical calculations, it will be convenient to modify this
formula and make it regular at $|t|=0$. We will use
\begin{equation}
\displaystyle \bar n_{1p} \simeq
\frac{\alpha}{2\pi}\ln^2\left(1+\frac{|t|}{\Lambda_\gamma^2}\right).
\label{ngamma}
\end{equation}
Within the double logarithmic accuracy, $\Lambda_\gamma$ preserve  its
meaning as a minimal frequency of the soft emission.

When estimating the average number of magnetic--bremsstrahlung photons
$\bar n_{(\mu_{an})}\equiv \bar n_{2\scN}$, the cut--off parameter
$\omega_{max(\mu_{an})}$ must be chosen in such a way that the expression
(\ref{sph}) turns into logarithm (the squares of the  logarithms must be
cancelled). This is achieved with $\omega_{max(\mu_{an})}\sim
\sqrt{m\varepsilon}$:

\begin{equation}
\di
\bar n_{2\scN} \simeq
\frac{\kappa\alpha}{2\pi}\ln\frac{|t|}{m^2},
\qquad \kappa=\ln\frac{m^2\sqrt{|t|}}{\omega_{min}^2\varepsilon}.
\label{sph2p}
\end{equation}

Thus, for $\omega_{max(\mu_{an})}\sim \sqrt{m\varepsilon}$ parameter
$\kappa$ is not logarithmically large, so the estimate (\ref{sph2p}) is in
agreement with the expression for $\sigma_\gamma^{(\mu_{an})}$, given in
(\ref{sitot}). Of course, the contribution to $\kappa$ is also given by
sub--leading logarithmic terms, which are not present in (\ref{sph}) and
can be found only in more accurate calculations. Further on, however, the
only fact important for our consideration is that the average number of
electric--dipole soft photons considerably exceeds that of
magnetic--bremsstrahlung ones.
\begin{equation}
\di \bar n_{2n}\sim \bar n_{2p} \ll \bar n_{1p},
\label{n12}
\end{equation}

\subsection{How soft photon emission can modify observables \label{DPamplitudes}}

Thus, beyond perturbation theory, the Dirac and Pauli form factors
are disparate, which is revealed in Eqs.~(\ref{disparity}),
(\ref{n12}). Therefore, {\it nonperturbative} probability of soft
emission cannot be represented as a single multiplier outside the
pure elastic Rosenbluth cross section. Since the process under
discussion is classic, the soft emission depends not only on the
initial and final electron states, but also on the peculiar
characteristics of the quasiclassical trajectory. The multiple
emission of soft photons at the alteration of electron trajectory
proceeds during the period of time $\tau$ that is significantly
longer that the inverse characteristic photon frequency,
$\omega\tau\gg 1$. In this situation, contrary to the one--photon
approximation, allowance must be made for the soft emission over
large angles.  The electron trajectory is formed by the
superposition of the electromagnetic fields, generated individually
by the Dirac and Pauli form factors, which leads to the two distinct
subsystems of the soft photon emission.

On the basis of the above qualitative considerations,  we put forward the
idea, that the Dirac and Pauli scattering amplitudes with emission of $n$
photons differ from the corresponding elastic amplitudes and can be
obtained by multiplying those on the two different {\it nonperturbative}
amplitudes of $n$ soft photon emission $A_{1{\scriptscriptstyle N}}(n)$,
$A_{2{\scriptscriptstyle N}}(n)$. Modification of the scattering
amplitudes leads to the modification of the form factors:
\[
\displaystyle F^{(n\gamma)}_{1{\scriptscriptstyle N}}
=A_{1{\scriptscriptstyle N}}(n)F_{1{\scriptscriptstyle N}}, \nonumber
\qquad
 F^{(n\gamma)}_{2{\scriptscriptstyle N}} = A_{2{\scriptscriptstyle
N}}(n)F_{2{\scriptscriptstyle N}}.
\]

As we discussed earlier, in the polarization transfer experiments
only events without real photon emission ($n=0$) can contribute. In
Rothenbluth experiments, on the contrary, the processes with
emission of any number $n$ of soft photons occur. Thus the {\it
nonperturbative} amplitudes of the soft photon emission are
different for these two experiments.

About the photon emission amplitudes $A_{1{\scriptscriptstyle
N}}(n)$, $A_{2{\scriptscriptstyle N}}(n)$ it is known, that their
averaged squares are equal to the corresponding probabilities
calculated via the Poisson distributions:
\begin{eqnarray}
\displaystyle \langle |A_{1{\scriptscriptstyle
N}}(n)|^2\rangle&=&w_{1{\scriptscriptstyle N}}(n)=\frac{\bar
n_{1{\scriptscriptstyle N}}^n}{n!}e^{-\bar n_{1{\scriptscriptstyle N}}},
\\
\langle |A_{2{\scriptscriptstyle N}}(n)|^2\rangle&=&
w_{2{\scriptscriptstyle N}}(n)=\frac{\bar n_{2{\scriptscriptstyle
N}}^n}{n!}e^{-\bar n_{2{\scriptscriptstyle N}}}.
\label{P}
\end{eqnarray}

The averaged numbers of the "dipole"\-type $\bar n_{1{\scriptscriptstyle
N}}$ and "magnetic--bremsstrahlung"\ type $\bar n_{2{\scriptscriptstyle
N}}$ photons are different and as well depend on the nucleon, in the field
of which the emission is formed (target nucleon in our case). For the
target proton, photon emission in the short-range Pauli field is
negligible in comparison with that in the long-range Dirac field, $\bar
n_{2p}\ll \bar n_{1p}$, $\bar n_{1p}$ will be give below in
Eq.(\ref{ngamma}). For the neutron target, $\bar n_{1n}\ll
\bar n_{1p}$ because the Dirac form factor is small, and $\bar n_{2n}\sim \bar
n_{2p}$. For our purposes it would be enough to estimate
\[
A_{1n}(0)\simeq A_{2n}(0)\simeq 1, \quad A_{1n}(n)\simeq A_{2n}(n)\simeq 0
\; \mbox{for} \; n\geqslant 1,
\]
Within this assumption, the electric and magnetic form factors of neutron
are not modified by the soft photon emission and are given by the standard
expressions
\begin{eqnarray}
G_{En}(t)&=&F_{1n}(t)+\frac{t}{4\mN^2}F_{2n}(t), \nonumber
\\
G_{Mn}(t)&=&F_{1n}(t)+ F_{2n}(t),
\label{Gneutron}
\end{eqnarray}
So, our model predict that the ratio of the neutron form factors, obtained
in different types of experiments must be approximately the same.

\subsection{Form factors from Rosenbluth scattering \label{FFRos}}

To obtain the proton form factors, let us separate from the
Rosenbluth formula the combination of the form factors
$\sigma_{ext}(t,\theta)$ which enters the differential cross
section

\[
\di
\frac{d\sigma_{Ros}}{d\Omega}=f_{rec}\sigma_{Mott}\sigma_{ext}(t,\theta),
\]
\begin{equation}
\begin{array}{r}
\di \sigma_{ext}(t,\theta)=
\frac{\left[G_{Ep}^{(Ros)}\right]^2-
\di\frac{t}{4M_\scN^2}\left[G_{Mp}^{(Ros)}\right]^2}{1-\di\frac{t}{4M_\scN^2}}
\\[5mm] \di
-\frac{t}{2M_\scN^2}\left[G_{Mp}^{(Ros)}\right]^2\tan^2\frac{\theta}{2}.
\end{array}
\label{Ros}
\end{equation}
 The squares of
the form factors are calculated as
\begin{equation}
\begin{array}{c}
\displaystyle \left[G^{(Ros)}_{Ep}\right]^2=
\sum_{n=0}^\infty\langle\left[A_{1p}(n)F_{1p}
                 +\frac{t}{4\mN^2}A_{2p}(n)F_{2p}\right]^2\rangle=
\\[5mm]
\di =\left[F_{1p}+\frac{t }{4\mN^2} F_{2p} \right]^2
\\[2mm] \di
-\left[1-\sum_{n=0}^{\infty}\langle
A_{1p}(n)A_{2p}(n)\rangle\right]\frac{t}{2\mN^2} F_{1p}F_{2p},
\\[5mm]
\displaystyle \left[G^{(Ros)}_{Mp}\right]^2=
\sum_{n=0}^\infty\langle\left[A_{1p}(n)F_{1p}+A_{2p}(n)F_{2p}\right]^2\rangle=
\\[5mm]
\di =\left[F_{1p}+F_{2p}\right]^2-
2\left[1-\sum_{n=0}^{\infty}\langle
A_{1p}(n)A_{2p}(n)\rangle\right]F_{1p}F_{2p}.
\end{array}
\label{Ros1}
\end{equation}
In Eqs.(\ref{Ros1}) there is an interference term of Dirac and
Pauli form factors, that is interference of the soft emission of
different types.
In the first summand of Eq. (\ref{Ros}), which gives the dominant
contribution to the scattering through small angles, the two
interference terms cancel out. Therefore the interference effects
would be noticeable only at large scattering angles.

At large $t$ the experimental procedure includes two steps. At the first
step the coefficient before the $\tan^2(\theta/2)$ is measured and
magnetic form factor $G_{Mp}^{(Ros)}(t)$ is deduced. Having it known and
provided that the accuracy of the measurement is high enough, the electric
form factor  $G_{Ep}^{(Ros)}(t)$ is extracted. According to this
procedure, the experimentally observed values are described by Eqs.
(\ref{Ros1}), which including interference terms.

The averaged interference terms in the two special cases are
\begin{eqnarray}
\di  \sum_{n=0}^{\infty}\langle A_{1p}(n)A_{2p}(n)\rangle=1
\quad
\mbox{if}
\quad &
1)\; \bar n_{1p} =0,\;\bar n_{2p}=0; \nonumber
\\
& 2)\;
\bar n_{1p}=\bar n_{2p}.
\label{AA}
\end{eqnarray}
The properties (\ref{P}), (\ref{AA}) can be satisfied by choosing
the amplitudes in a very simple form
\begin{equation}
A_{1p}(n) =\sqrt{w_{1p}(n)}
\qquad
A_{2p}(n) =\sqrt{w_{2p}(n)}
\label{16}
\end{equation}

Substituting (\ref{16}) into (\ref{Ros1}) yields expressions,
which can be further simplified for  $\bar n_{2p}\ll \bar n_{1p}$:

\begin{equation}
\begin{array}{c}
\di  \left[G^{(Ros)}_{Ep}\right]^2
 =\left[F_{1p}+\frac{t}{4\mN^2}F_{2p}\right]^2
\\[4mm] \di
 -\left[1-e^{-(\sqrt{\bar n_{1p}}-\sqrt{\bar n_{2p}})^2/2}\right]\frac{t}{2M_\scN^2}F_{1p}F_{2p}\simeq
\\[3mm]
\di \simeq \left[F_{1p}+\frac{t}{4\mN^2}F_{2p}\right]^2-
\left(1-e^{-\bar n_{1p}/2}\right)\frac{t}{2\mN^2}F_{1p}F_{2p},
\\[6mm]
\di
\left[G^{(Ros)}_{Mp}\right]^2=\left[F_{1p}+F_{2p}\right]^2
\\[4mm] \di
-2\left[1-e^{-(\sqrt{\bar n_{1p}}-\sqrt{\bar
n_{2p}})^2/2}\right]F_{1p}F_{2p}\simeq
\\[3mm]
\di \simeq \left[F_{1p}+F_{2p}\right]^2-2\left( 1-e^{-\bar
n_{1p}/2}\right)F_{1p}F_{2p}.
\end{array}
\label{GRos}
\end{equation}

\begin{equation}
R^{(Ros)}_{p}=\sqrt{\frac{\left[G^{(Ros)}_{Ep}\right]^2}{\left[G^{(Ros)}_{Mp}\right]^2}}
\label{RRos}
\end{equation}

The average number of soft photons is given in Eq.~(\ref{ngamma}).
When fitting the experimental data, the minimal frequency
$\Lambda_\gamma$ of the soft emission, according to its physical
status, is considered as a free fit parameter.

As one could see the asymmetry in the soft photon emission by Dirac and
Pauli form factors considerably modifies (with respect to naive
expectations) the relations between the theoretically calculated
$F_{1,2p}$ and experimentally measured $G^{(Ros)}_{E,Mp}$ quantities . One
should however remember, that this formulas are only applicable to the
Rosenbluth extraction.

\subsection{Form factors from recoil polarization  experiments \label{FFpol}}

In elastic scattering of polarized electrons from a nucleon, the nucleon
is transferred a polarization, whose components along ($P_l$) and
perprerdiclular to ($P_t$)  the nucleon momentum  are proportional to
$G_M^2$ and $G_EG_M$, respectively. Such polarization transfer can only
proceed without the soft photon emission. Thus, the the Dirac form factor
is multiplied by the probability amplitude with zero photons
$A_{1p}(0)=e^{-\bar n_{1p}/2}$ (only one nonzero term remains in the whole
sum). The modification of Pauli form factor is negligible (within the
accuracy of our consideration).


As a result, the experimentally measured ratio $P_t/P_l\sim G_E G_M/
G_M^2=G_E/ G_M$ is proportional to the ratio of the form factors, which
can be calculated as

\begin{equation}
 \displaystyle
R^{(pol)}_p\equiv
\frac{\mu_pG^{(pol)}_{Ep}}{G^{(pol)}_{Mp}}=
\frac{\mu_p\left(e^{-\bar
n_{1p}/2}F_{1p}+\displaystyle\frac{t}{4M_{\scriptscriptstyle
N}^2}F_{2p}\right)}{e^{-\bar n_{1p}/2}F_{1p}+F_{2p}}.
\label{Rpol}
\end{equation}
with the Dirac and Pauli form factors, defined differently from the naive
expectation as well as from (\ref{RRos}).

\subsection{Form factors from beam--target asymmetry \label{FFbta}}

Another type of experiments is scattering of a polarized electron
against a polarized target nucleon.  With the target polarization
axis in the scattering plane and perpendicular to the momentum
transfer $\vec q$, the asymmetry $A_{TL}$ (the difference of the
cross sections for opposite electron beam helicities) is
proportional to $G_E G_M$. With the target polarization axis in the
scattering plane and parallel to $\vec q$, the asymmetry $A_{T}$ is
proportional to $G_M^2$. Experimentally measured is the ratio: \begin{equation}
\frac{A_{TL}}{A_{T}}\sim\frac{G_E G_M}{G_M^2} \ . \end{equation} In such
experiments the final state is inclusive, so the soft photon
emission is possible, like in Rosenbluth scattering. The value $G_E
G_M$ must be calculated as follows \begin{equation}
\begin{array}{c}
\di  \left[G_{Ep}G_{Mp}\right]^{(bta)}
=\sum_{n=0}^\infty\langle
  \left[A_{1p}(n)F_{1p}+A_{2p}(n)F_{2p}\right]\times
\\[2mm] \di
  \times \left[A_{1p}(n)F_{1p}+\frac{t}{4\mN^2}A_{2p}(n)F_{2p}\right]
                  \rangle=
\\[3mm]  \di
=  \left[F_{1p}+\frac{t}{4\mN^2}A_{2p}(n)F_{2p}\right]
  \left[F_{1p}+F_{2p}\right]
\\[2mm] \di
-\left(1-e^{-\bar n_{1p}/2}\right)F_{1p}F_{2p},
\end{array}
\label{GEGMbta}
\end{equation}
and $G_M^2$ is the same as in the Rosenbluth experiments
$[G_{Mp}^2]^{bta}=[G_{Mp}^2]^{Ros}$. So, the ratio of the form factors
\begin{equation}
 \displaystyle
R^{(bta)}_p\equiv
\frac{\mu_p[G_{Ep}G_{Mp}]^{(bta)}}{[G_{Mp}^2]^{(bta)}} \
\label{Rbta}
\end{equation}
is defined differently from (\ref{Rpol}) as well as from
(\ref{RRos}).

\subsection{Recap}

Let us shortly summarize the results if this section. To interpret
experimental data from different kinds of experiments, different
expressions for the form factors must be used.
All the data can be physically and mathematically described with the
same theoretically defined form factors $F_{1,2p}$ (\ref{10}) and
the same average number of soft electric--dipole photons $\bar
n_{1p}$ . The electric and magnetic form factors and their ratios
are given, however, by different expressions: Eq. (\ref{GRos}) for
Rosenbluth separation technique, Eq. (\ref{Rpol}) for polarization
transfer, and Eq. (\ref{Rbta}) for beam--target asymmetry
experiments.

\section{Fit of the experimental data \label{fit}}

The model described in previous sections is used to fit all
experimental data  on the form factors (with the references
summarized in Tables~\ref{tab:FFp},\ref{tab:FFn}) available to date.
In the framework of this model the form factors $F^{p,n}_{1,2}$ are
described by Eqs. (\ref{10}) with the four phenomenological
logarithmic functions $q_{s,v}$, $\mu_{s,v}$ given in (\ref{qmu}).
The pole positions $m_{\rho,\omega,\phi}$ were taken as masses of
vector mesons from PDG\cite{Yao:2006px}. Parameters $p_1$, $p_2$ in
these functions are calculated according to (\ref{pQCD}) with the
effective number of quark flavors being given by interpolation
formula \begin{equation} \displaystyle
 n_f=2+\frac{|t|}{|t|+4m_s^2}+\frac{|t|}{|t|+4m_c^2},
\label{nf}
\end{equation}
where $m_s\approx 0.15\GeV$ and $m_c\approx 1.5 \GeV$ are masses of
strange and charmed quarks correspondingly. The form factors in Rosenbluth
experiments and the ratio in polarization experiments are determined in
Eqs. (\ref{GRos}), (\ref{Rpol}) with the averaged number of the emitted
soft photons from (\ref{ngamma}).

\begin{table}
\begin{center}
\caption{Data on proton form factors. \label{tab:FFp}}
\tabcolsep6.8pt
\renewcommand{\arraystretch}{1.3}
\begin{tabular}{||l|l|l||}
\hline \hline
Measurement  & $Q^2$-range     & reference                      \\
\hline
$G_{Ep}$    &                 &                                 \\
$p(e,e')$    & ~0.01 - ~0.05   & Simon et al.~\cite{Simon:1980hu}       \\
             & ~0.04 - ~1.75   & Price et al.~\cite{Price:1971zk}       \\
             & ~0.39 - ~1.95   & Berger et al.~\cite{Berger:1971kr}      \\
             &  ~1.00 - ~3.00  & Walker et al.~\cite{Walker:1993vj}      \\
             &  1.75 - ~8.83   & Andivahis et al.~\cite{Andivahis:1994rq}   \\
$p(e,p')$    &  2.64 - 4.10    & Qattan et al.~\cite{Qattan:2004ht}   \\
             &  0.65 - 5.2     & Christy et al.~\cite{Christy:2004rc}   \\
$d(e,e'p)$   & ~0.27 - ~1.76   & Hanson et al~\cite{Hanson:1973vf}       \\

\hline
$G_{Mp}$     &                 &                                 \\
$p(e,e')$    &  ~0.02 - ~0.15  & Hoehler et al.~\cite{Hohler:1976ax}     \\
             &  ~0.16 - ~0.86  & Janssens et al.~\cite{Janssens:1966}    \\
             &  ~0.39 - ~1.75  & Berger et al.~\cite{Berger:1971kr}      \\
             &  ~0.67 - ~3.00  & Bartel et al.~\cite{Bartel:1973rf}      \\
             &  ~1.00 - ~3.00  & Walker et al.~\cite{Walker:1993vj}      \\
             &  ~1.50 - ~3.75  & Litt et al.~\cite{Litt:1969my}        \\
             &  ~1.75 - ~7.00  & Andivahis et al.~\cite{Andivahis:1994rq}   \\
             &  ~2.86 - 31.2   & Sill et al.~\cite{Sill:1992qw}        \\
$d(e,e'p)$   &  ~0.27 - ~1.76  & Hanson et al~\cite{Hanson:1973vf}       \\
\hline
$G_{Ep}/G_{Mp}$     &                 &                                 \\
$p(\stackrel{\rightarrow}{e},e'\stackrel{\rightarrow}{p})$
             & ~0.37 - ~0.44   & Pospischil et al.~\cite{Pospischil:2001pp}  \\
             & ~0.38 - ~0.50   & Milbrath et al.~\cite{Milbrath:1997de}    \\
             & ~0.40           & Dieterich et al.~\cite{Dieterich:2000mu}   \\
             & ~0.49 - ~3.47   & Jones et al.~\cite{Jones:1999rz}       \\
             & 1.13            & Mac Lachlan et al.~\cite{MacLachlan:2006vw}   \\
             & ~3.50 - ~5.54   & Gayou et al.~\cite{Gayou:2001qd,Punjabi:2005wq}       \\
 $\stackrel{\rightarrow}{p}(\stackrel{\rightarrow}{e}, e'p)$
             & 0.15 - 0.65     & Crawford et al.~\cite{Crawford:2006rz}       \\
 $\stackrel{\rightarrow}{p}(\stackrel{\rightarrow}{e}, e'p)$
             & 1.51            & Jones et al.~\cite{Jones:2006kf}       \\
\hline
\end{tabular}
\end{center}
\end{table}

\begin{table}
\begin{center}
\caption{Data on neutron form factors. \label{tab:FFn}}
\tabcolsep6.8pt
\renewcommand{\arraystretch}{1.3}
\begin{tabular}{||l|l|l||}
\hline \hline
Measurement  & $Q^2$-range     & reference                      \\
\hline
$G_{En}$    &                &                                  \\
$\stackrel{\rightarrow}{d}(\stackrel{\rightarrow}{e},e'n)p$
             & 0.21           & Passchier et al.~\cite{Passchier:1999cj}      \\
             & 0.50           & Zhu et al.~\cite{Zhu:2001md}          \\
$\overrightarrow{^3He}(\stackrel{\rightarrow}{e},e'n)$
             & 0.40           & Becker et al.~\cite{Becker:1999tw,Bermuth:2001mu,Golak:2000nt}\\
$d(\stackrel{\rightarrow}{e},e'\stackrel{\rightarrow}{n})p$
             & 0.5, $\;$ 1.0    & Warren et al. \cite{Warren:2003ma,Savvinov:2004gt}\\
             & 0.67           & Rohe et al.~\cite{Rohe:1999sh}    \\
             & 0.67           & Bermuth et al.~\cite{Bermuth:2003qh} \\
           ? &                & Schiavilla et al.~\cite{Schiavilla:2001qe}\\
$d(e,e')$    & 0.27 - 1.76  & Hanson et al~\cite{Hanson:1973vf}        \\
             & 1.75 - 4.00  & Lung et al.~\cite{Lung:1992bu}         \\
\hline
$G_{Mn}$     &                &                                  \\
$d(e,e'n)p$  & ~0.07 - ~0.89  & Kubon et al.~\cite{Kubon:2001rj}        \\
             & ~0.11          & Anklin et al.~\cite{Anklin:1994ae}       \\
             & 0.11 - ~0.26 & Markowitz et al.~\cite{Markowitz:1993hx}    \\
             & 0.13 - ~0.61 & Bruins et al.~\cite{Bruins:1995ns}       \\
             & ~0.24 - ~0.78  & Anklin et al.~\cite{Anklin:1998ae}       \\
$d(e,e'p)$   & 0.27 - ~1.76 & Hanson et al~\cite{Hanson:1973vf}        \\
$\overrightarrow{^3He}(\stackrel{\rightarrow}{e},e'n)$
             & 0.10, $\;$ 0.20  & Xu et al.~\cite{Xu:2000xw,Golak:2000nt}            \\
             &  0.19          & Gao et al.~\cite{Gao:1994ud}           \\
             & ~0.3 - ~0.6    & Xu et al.~\cite{Xu:2002xc}            \\
$d(e,e')$    & ~1.75 - ~4.00  & Lung et al.~\cite{Lung:1992bu}         \\
             & ~2.50 - 10.0   & Rock et al.~\cite{Rock:1982gf}         \\
\hline
$G_{En}/G_{Mn}$  &                &                                  \\
 $d(\stackrel{\rightarrow}{e},e'\stackrel{\rightarrow}{n})p$
             & 0.30 - 0.8     & Glazier et al.~\cite{Glazier:2004ny}      \\
             & 0.15           & Herberg et al.~\cite{Herberg:1999ud}       \\
             & 0.26           & Eden et al.~\cite{Eden:1994ji}          \\
             & 0.34           & Ostrick et al.~\cite{Ostrick:1999xa}       \\
             & 0.49 - ~1.47   & Madey et al.\cite{Madey:2003av,Reichelt:2003iw,Plaster:2005cx}   \\
\hline
\end{tabular}
\end{center}
\end{table}

In the model under consideration the coefficients of the multipole
expansions obey the  SR (\ref{SR}) and, thus, not all of them are
free fit parameters. In the isovector sector from $N_v$ parameters
$\tilde a_{v(k)}$, which satisfy two SR, and $N_v$ parameters
$\tilde b_{v(k)}$, which satisfy three SR, $2N_v-5$ parameters are
independent. In the isoscalar sector one ends up analogically with
$2N_s-5$ independent free parameters.

The model contains 10 more parameters: 9 in the logarithmic
functions (\ref{qmu}) and 1 in Eq. (\ref{ngamma}) for the average
number of soft photons. The status of $g_{1,2{s,v}}$ and
$h_{1,2(s,v)}$ was discussed in subsection \ref{logarithcorr},
$\Lambda$ approximately equals  the QCD scale and $\Lambda_\gamma$
is a minimal frequency of soft photon emission.

As we have already mentioned, to satisfy three of the introduced SR,
at least three vector meson families are to be considered. Below we
describe the fit obtained in the framework of the simplest
modification of our model with $N_v=3$ and the isovector poles
identified with the squared $\rho-$meson masses. In this model all
the three parameters $\tilde b_{v(k)}\equiv \tilde b_{\rho(k)}$ are
fixed by the SR and not fitted. Three parameters $\tilde
a_{v(k)}\equiv \tilde a_{\rho(k)}$ satisfy two SR, so only one of
them is independent. It is chosen as $\kappa_{1v}$, introduced in
(\ref{kappa}).

In this simplest modification only three $\omega$ mesons are taken
into account in the isoscalar sector and $\phi$-mesons are
neglected. Such simplification is physically justified because the
$eN$ interaction through the $\phi$ mesons contributes only little
to the structure of the form factors. Indeed, for the ideal
singlet--octet mixing the $\phi$ mesons consist of two strange
quarks $\phi=\bar ss$, and thus interact only with the virtual
strange component of the nucleon, which is insignificant for the
nucleon structure. Deviations from the ideal mixing are suppressed
by small parameters. To the zeroth order on these parameters,
interactions of $\phi$ mesons with $ud-$ component of the nucleon
are negligible. After excluding  $\phi$ mesons from the isoscalar
sector of the model, parameters $\tilde b_{s(k)}=\tilde
b_{\omega(k)}$ are fixed by the three SR and among three parameters
$\tilde a_{s(k)}=\tilde a_{\omega(k)}$ only one is independent. It
is chosen as $\kappa_{1s}$ from Eq.~(\ref{kappa}).

This simplest modification of the model without $\phi$ mesons is
used for the preliminary joined fit of the Rosenbluth and
polarization data. There are 12 fit parameters common for the 5 sets
of data. The result of the fit is shown in
Figs.~\ref{fig:Gprot}--\ref{fig:Rneut}, the accuracy of the fit
being $\chi^2/dof=0.86$. The parameters of the model, obtained in
our fit are:

\begin{equation}
\begin{array}{c} \di
\kappa_{1v}=-0.896, \quad \kappa_{1s}=-0.0814,
\\ \di
g_{1v}=0.0949, \quad h_{1v}=0,
\\ \di
g_{1s}=0.0138, \quad h_{1s}=0,
\\ \di
g_{2v}=0.326, \quad h_{2v}=-0.118,
\\ \di
g_{2s}=0.0813, \quad h_{2s}=-0.568
\\ \di
\Lambda=0.163 \GeV, \Lambda_\gamma=0.00405 \GeV
\end{array}
\label{fitres-3rho3omega}
\end{equation}
The errors are hard to estimate due to nonlinearity of the problem.

An attempt to improve this fit by including $\phi$-mesons (of all
the three generations considered here) leads to a better accuracy of
the fit, $\chi^2/dof=0.81$. Keeping in mind, however, that there are
additional 6 fit parameters for this case (three $\tilde a_{s}$ and
three $\tilde b_s$), this improvement of accuracy does not look
significant. In any case, the difference between the curves for the
cases with and without $\phi-$mesons is hardly noticeable. For this
reason we prefer to leave the detailed discussion of the
$\phi-$meson contribution for further investigation.

\section{Predictions \label{predictions}}

In this section we shortly summarize the predictions of our model, which
in principle can be tested experimentally.

\begin{enumerate}
\item For proton the formulas for experimentally measured electric and
magnetic form factors and their ratios are different for Rosenbluth
(see Eq.~(\ref{GRos}), (\ref{RRos})), polarization (see
Eqs.~(\ref{Rpol})) and beam--target asymmetry (see Eq.~(\ref{Rbta}))
experiments. The ratios are shown in Fig.~\ref{fig:Rprot}. They
satisfy the inequality
\[
R^{(pol)}(Q^2)<R^{(bta)}(Q^2)<R^{(Ros)}(Q^2)
\]

\item Rothenbluth electric form factor of the proton does not goes to
zero, the ratio of the form factors grows with increasing $Q^2$. The
ratio, measured in the polarization transfer experiments, crosses zero at
$Q^2\sim 9-10 \GeV^2$. The ratio, measured in the beam-target asymmetry
experiments, crosses zero at $Q^2\sim 17-19 \GeV^2$

\item  For neutron the data from all three types of experiments coincide:
\begin{eqnarray} \nonumber
\di G_{En}^{(Ros)}=G_{En}^{(pol)}=G_{En}^{(bta)}=G_{En}
\\[3mm] \di  \nonumber
G_{Mn}^{(Ros)}=G_{Mn}^{(pol)}=G_{Mn}^{(bta)}=G_{Mn}
\\[3mm] \di  \nonumber
R_{n}^{(Ros)}=R_{n}^{(Ros)}=R_{n}^{(Ros)}=G_{En}/G_{Mn}
\end{eqnarray}
and are given by conventional formulas (\ref{Gneutron}). The ratio of the
form factors is shown in Fig.~\ref{fig:Rneut}.

\end{enumerate}

\section{Conclusions \label{conclusions}}

Thus, we propose an interpretation of the electromagnetic nucleon
form factors on the basis of the following physical conceptions:

\begin{enumerate}
\item multipole structure of the form factors, which correspond to the
vector meson dominance with taking into account at least three
$\omega\rho\phi-$ families.

\item logarithmic dependencies of the coefficients of the multipole
expansions reflect the renormalization of the nuceleon--meson coupling
constants by the quark--gluon effects at small distances;

\item Superconvergence Relations for the meson parameters ensures the
agreement of the multipole form factors asymptotics the pQCD asymptotics;
in this way we account for QCD effects at the level of valence quarks.

\item form factors, extracted from the cross sections, are modified by the
effect of the soft emission, which is different for different experiments.
This effect allows to resolve the discrepancy between the Rosenbluth and
polarization measurements.
\end{enumerate}

In the framework of this model the Dirac $F_{1p,n}$ and Pauli
$F_{2p,n}$ form factors are the same for all types of experiments.
They are obtained from the fit with the results explicitly given in
the following equations:
\begin{equation}
\begin{array}{c} \displaystyle
F_{1p,n}=\frac12\left[ F_{1s}\pm F_{1v} \right],
\qquad
F_{2p,n}=\frac12\left[ F_{2s}\pm F_{2v} \right], 
\\[5mm] \displaystyle
F_{1s}=
\left[
\frac{0.923}{0.612+Q^2}+\frac{-1.314}{2.031+Q^2}+\frac{0.391}{2.789+Q^2}
\right]  \times
\\ \displaystyle
\left[1+0.0138\ln^2\left(1+\frac{Q^2}{0.0266}\right)\right]^{-p_1}
\\[5mm] \displaystyle
F_{1v}=\left[\frac{-0.793}{0.602+Q^2}+\frac{16.018}{2.147+Q^2}+\frac{-15.226}{2.958+Q^2}
\right]  \times
\\ \displaystyle
\left[1+0.0949\ln^2\left(1+\frac{Q^2}{0.0266}\right)\right]^{-p_1}
\\[5mm] \displaystyle
F_{2s}=
\left[
\frac{-0.135}{0.612+Q^2}+\frac{0.388}{2.031+Q^2}+\frac{-0.253}{2.789+Q^2}
\right]  \times
\\ \displaystyle
\left[1-0.568\ln\left(1+\frac{Q^2}{0.0266}\right)+0.0813\ln^2\left(1+\frac{Q^2}{0.0266}\right)\right]^{-p_2}
\\[5mm]  \displaystyle
F_{2v}=\left[\frac{3.893}{0.602+Q^2}+\frac{-11.295}{2.147+Q^2}+\frac{7.402}{2.958+Q^2}
\right]   \times
\\ \displaystyle
\left[1-0.118\ln\left(1+\frac{Q^2}{0.0266}\right)+0.326\ln^2\left(1+\frac{Q^2}{0.0266}\right)\right]^{-p_2}
\\[5mm] \displaystyle
p_1=2+\frac{3.555}{11-0.67n_f}, \quad p_2=\frac{2.667}{11-0.67n_f},
\\ \displaystyle
n_f=2+\frac{Q^2}{Q^2+0.0900}+\frac{Q^2}{Q^2+9.00}
\end{array}
\end{equation}

The experimentally measured form factors and their ratios are shown
in Figs.~\ref{fig:Gprot},  \ref{fig:Rprot}, \ref{fig:Gneut},
\ref{fig:Rneut}. They are given in Eqs.~(\ref{Gneutron}) for neutron
and Eqs.~(\ref{GRos}), (\ref{RRos}), (\ref{Rpol}) (\ref{Rbta}) for
proton. They depend not only on Dirac and Pauli form factors, but
also on the average number of the soft photons $\bar{n}_{1p}$,
emitted in the field of the Dirac proton form factor in the
experiments with inclusive final states \begin{equation}
\bar{n}_{1p}=0.116\cdot10^{-2}\ln^2\left(1+\frac{Q^2}{0.164\cdot
10^{-4} \GeV^2 }\right)\ . \label{barn1p_} \end{equation}

\begin{acknowledgement}
GV is grateful to the participants of the Theory Group Seminar in
JINR, Dubna, where he had a pleasure to present his talk, for their
comments and suggestions. Discussions with Prof. A. Efremov were
especially helpful.
\end{acknowledgement}

\bibliographystyle{epj} 
\bibliography{FFreferences}

\begin{thebibliography}{91}

\bibitem{Day:2007ed}
D.~Day, Eur. Phys. J. \textbf{A31}, 560 (2007)

\bibitem{Friedrich:2003iz}
J.~Friedrich, T.~Walcher, Eur. Phys. J. \textbf{A17}, 607 (2003),
  \texttt{hep-ph/0303054}

\bibitem{Tomasi-Gustafsson:2005kc}
E.~Tomasi-Gustafsson, F.~Lacroix, C.~Duterte, G.I. Gakh, Eur. Phys. J.
  \textbf{A24}, 419 (2005), \texttt{nucl-th/0503001}

\bibitem{Arrington:2002cr}
J.~Arrington, Eur. Phys. J. \textbf{A17}, 311 (2003), \texttt{hep-ph/0209243}

\bibitem{Bodek:2003ed}
A.~Bodek, H.~Budd, J.~Arrington, AIP Conf. Proc. \textbf{698}, 148 (2004),
  \texttt{hep-ex/0309024}

\bibitem{Budd:2003wb}
H.~Budd, A.~Bodek, J.~Arrington (2003), \texttt{hep-ex/0308005}

\bibitem{Kelly:2004hm}
J.J. Kelly, Phys. Rev. \textbf{C70}, 068202 (2004)

\bibitem{Punjabi:2005wq}
V.~Punjabi et~al., Phys. Rev. \textbf{C71}, 055202 (2005),
  \texttt{nucl-ex/0501018}

\bibitem{Brodsky:1974vy}
S.J. Brodsky, G.R. Farrar, Phys. Rev. \textbf{D11}, 1309 (1975)

\bibitem{Lepage:1980fj}
G.P. Lepage, S.J. Brodsky, Phys. Rev. \textbf{D22}, 2157 (1980)

\bibitem{Belitsky:2002kj}
A.V. Belitsky, X.d. Ji, F.~Yuan, Phys. Rev. Lett. \textbf{91}, 092003 (2003),
  \texttt{hep-ph/0212351}

\bibitem{Brodsky:2003pw}
S.J. Brodsky, J.R. Hiller, D.S. Hwang, V.A. Karmanov, Phys. Rev. \textbf{D69},
  076001 (2004), \texttt{hep-ph/0311218}

\bibitem{Guidal:2004nd}
M.~Guidal, M.V. Polyakov, A.V. Radyushkin, M.~Vanderhaeghen, Phys. Rev.
  \textbf{D72}, 054013 (2005), \texttt{hep-ph/0410251}

\bibitem{Gari:1992tq}
M.F. Gari, W.~Kruempelmann, Phys. Rev. \textbf{D45}, 1817 (1992)

\bibitem{Lomon:2001ga}
E.L. Lomon, Phys. Rev. \textbf{C64}, 035204 (2001), \texttt{nucl-th/0104039}

\bibitem{Lomon:2002jx}
E.L. Lomon, Phys. Rev. \textbf{C66}, 045501 (2002), \texttt{nucl-th/0203081}

\bibitem{Iachello:2004ki}
F.~Iachello, Eur. Phys. J. \textbf{A19}, Suppl129 (2004)

\bibitem{Bijker:2004yu}
R.~Bijker, F.~Iachello, Phys. Rev. \textbf{C69}, 068201 (2004),
  \texttt{nucl-th/0405028}

\bibitem{Mergell:1995bf}
P.~Mergell, U.G. Meissner, D.~Drechsel, Nucl. Phys. \textbf{A596}, 367 (1996),
  \texttt{hep-ph/9506375}

\bibitem{Hammer:1996kx}
H.W. Hammer, U.G. Meissner, D.~Drechsel, Phys. Lett. \textbf{B385}, 343 (1996),
  \texttt{hep-ph/9604294}

\bibitem{Hammer:2003ai}
H.W. Hammer, U.G. Meissner, Eur. Phys. J. \textbf{A20}, 469 (2004),
  \texttt{hep-ph/0312081}

\bibitem{Clementel:1956}
E.~Clementel, C.~Villi, Nuovo Cim. \textbf{4}, 1207 (1956)

\bibitem{Bergia:1961}
S.~Bergia et~al., Phys. Rev. Lett. \textbf{6}, 367 (1961)

\bibitem{Maximon:2000hm}
L.C. Maximon, J.A. Tjon, Phys. Rev. \textbf{C62}, 054320 (2000),
  \texttt{nucl-th/0002058}

\bibitem{Blunden:2003sp}
P.G. Blunden, W.~Melnitchouk, J.A. Tjon, Phys. Rev. Lett. \textbf{91}, 142304
  (2003), \texttt{nucl-th/0306076}

\bibitem{Blunden:2005ew}
P.G. Blunden, W.~Melnitchouk, J.A. Tjon, Phys. Rev. \textbf{C72}, 034612
  (2005), \texttt{nucl-th/0506039}

\bibitem{Arrington:2004ae}
J.~Arrington, Phys. Rev. \textbf{C71}, 015202 (2005), \texttt{hep-ph/0408261}

\bibitem{Bystritskiy:2007hw}
Y.M. Bystritskiy, E.A. Kuraev, E.~Tomasi-Gustafsson, Phys. Rev. \textbf{C75},
  015207 (2007)

\bibitem{Tvaskis:2005ex}
V.~Tvaskis et~al., Phys. Rev. \textbf{C73}, 025206 (2006),
  \texttt{nucl-ex/0511021}

\bibitem{Gould:1979pw}
R.J. Gould, Astrophys. J. \textbf{230}, 967 (1979)

\bibitem{Matveev:1973ra}
V.A. Matveev, R.M. Muradian, A.N. Tavkhelidze, Nuovo Cim. Lett. \textbf{7}, 719
  (1973)

\bibitem{Adamuscin:2002ca}
C.~Adamuscin, A.Z. Dubnickova, S.~Dubnicka, R.~Pekarik, P.~Weisenpacher, Eur.
  Phys. J. \textbf{C28}, 115 (2003), \texttt{hep-ph/0203175}

\bibitem{Hammer:2003qv}
H.W. Hammer, D.~Drechsel, U.G. Meissner, Phys. Lett. \textbf{B586}, 291 (2004),
  \texttt{hep-ph/0310240}

\bibitem{Belushkin:2005ds}
M.A. Belushkin, H.W. Hammer, U.G. Meissner, Phys. Lett. \textbf{B633}, 507
  (2006), \texttt{hep-ph/0510382}

\bibitem{Belushkin:2006qa}
M.A. Belushkin, H.W. Hammer, U.G. Meissner, U.G. Meissner (2006),
  \texttt{hep-ph/0608337}

\bibitem{Bass:2004xa}
S.D. Bass, Rev. Mod. Phys. \textbf{77}, 1257 (2005), \texttt{hep-ph/0411005}

\bibitem{Bass:2006cr}
S.D. Bass, C.A. Aidala, Int. J. Mod. Phys. \textbf{A21}, 4407 (2006),
  \texttt{hep-ph/0606269}

\bibitem{Carvalho:2005nk}
F.~Carvalho, F.S. Navarra, M.~Nielsen, Phys. Rev. \textbf{C72}, 068202 (2005),
  \texttt{nucl-th/0509042}

\bibitem{vonHarrach:2005at}
D.~von Harrach, Prog. Part. Nucl. Phys. \textbf{55}, 308 (2005)

\bibitem{Thomas:2006jf}
A.W. Thomas, R.D. Young, Nucl. Phys. \textbf{A782}, 1 (2007)

\bibitem{Rekalo:2003xa}
M.P. Rekalo, E.~Tomasi-Gustafsson, Eur. Phys. J. \textbf{A22}, 331 (2004),
  \texttt{nucl-th/0307066}

\bibitem{Berestetsky:1982aq}
V.b. Berestetsky, E.m. Lifshitz, L.p. Pitaevsky (????), oxford, Uk: Pergamon (
  1982) 652 P. ( Course Of Theoretical Physics, 4)

\bibitem{Yao:2006px}
W.M. Yao et~al. (Particle Data Group), J. Phys. \textbf{G33}, 1 (2006)

\bibitem{Simon:1980hu}
G.G. Simon, C.~Schmitt, F.~Borkowski, V.H. Walther, Nucl. Phys. \textbf{A333},
  381 (1980)

\bibitem{Price:1971zk}
L.E. Price et~al., Phys. Rev. \textbf{D4}, 45 (1971)

\bibitem{Berger:1971kr}
C.~Berger, V.~Burkert, G.~Knop, B.~Langenbeck, K.~Rith, Phys. Lett.
  \textbf{B35}, 87 (1971)

\bibitem{Walker:1993vj}
R.C. Walker et~al., Phys. Rev. \textbf{D49}, 5671 (1994)

\bibitem{Andivahis:1994rq}
L.~Andivahis et~al., Phys. Rev. \textbf{D50}, 5491 (1994)

\bibitem{Qattan:2004ht}
I.A. Qattan et~al., Phys. Rev. Lett. \textbf{94}, 142301 (2005),
  \texttt{nucl-ex/0410010}

\bibitem{Christy:2004rc}
M.E. Christy et~al. (E94110), Phys. Rev. \textbf{C70}, 015206 (2004),
  \texttt{nucl-ex/0401030}

\bibitem{Hanson:1973vf}
K.M. Hanson et~al., Phys. Rev. \textbf{D8}, 753 (1973)

\bibitem{Hohler:1976ax}
G.~Hohler et~al., Nucl. Phys. \textbf{B114}, 505 (1976)

\bibitem{Janssens:1966}
T.~Janssens et~al., Phys. Rev. \textbf{142}, 922 (1966)

\bibitem{Bartel:1973rf}
W.~Bartel et~al., Nucl. Phys. \textbf{B58}, 429 (1973)

\bibitem{Litt:1969my}
J.~Litt et~al., Phys. Lett. \textbf{B31}, 40 (1970)

\bibitem{Sill:1992qw}
A.F. Sill et~al., Phys. Rev. \textbf{D48}, 29 (1993)

\bibitem{Pospischil:2001pp}
T.~Pospischil et~al. (A1), Eur. Phys. J. \textbf{A12}, 125 (2001)

\bibitem{Milbrath:1997de}
B.D. Milbrath et~al. (Bates FPP), Phys. Rev. Lett. \textbf{80}, 452 (1998),
  \texttt{nucl-ex/9712006}

\bibitem{Dieterich:2000mu}
S.~Dieterich et~al., Phys. Lett. \textbf{B500}, 47 (2001),
  \texttt{nucl-ex/0011008}

\bibitem{Jones:1999rz}
M.K. Jones et~al. (Jefferson Lab Hall A), Phys. Rev. Lett. \textbf{84}, 1398
  (2000), \texttt{nucl-ex/9910005}

\bibitem{MacLachlan:2006vw}
G.~MacLachlan et~al., Nucl. Phys. \textbf{A764}, 261 (2006)

\bibitem{Gayou:2001qd}
O.~Gayou et~al. (Jefferson Lab Hall A), Phys. Rev. Lett. \textbf{88}, 092301
  (2002), \texttt{nucl-ex/0111010}

\bibitem{Crawford:2006rz}
C.B. Crawford et~al. (2006), \texttt{nucl-ex/0609007}

\bibitem{Jones:2006kf}
M.K. Jones et~al. (Jefferson Lab Resonance Spin Structure), Phys. Rev.
  \textbf{C74}, 035201 (2006), \texttt{nucl-ex/0606015}

\bibitem{Passchier:1999cj}
I.~Passchier et~al., Phys. Rev. Lett. \textbf{82}, 4988 (1999),
  \texttt{nucl-ex/9907012}

\bibitem{Zhu:2001md}
H.~Zhu et~al. (E93026), Phys. Rev. Lett. \textbf{87}, 081801 (2001),
  \texttt{nucl-ex/0105001}

\bibitem{Becker:1999tw}
J.~Becker et~al., Eur. Phys. J. \textbf{A6}, 329 (1999)

\bibitem{Bermuth:2001mu}
J.~Bermuth (A1), AIP Conf. Proc. \textbf{603}, 331 (2001)

\bibitem{Golak:2000nt}
J.~Golak, G.~Ziemer, H.~Kamada, H.~Witala, W.~Gloeckle, Phys. Rev.
  \textbf{C63}, 034006 (2001), \texttt{nucl-th/0008008}

\bibitem{Warren:2003ma}
G.~Warren et~al. (Jefferson Lab E93-026), Phys. Rev. Lett. \textbf{92}, 042301
  (2004), \texttt{nucl-ex/0308021}

\bibitem{Savvinov:2004gt}
N.~Savvinov (JLab E93-026), Braz. J. Phys. \textbf{34}, 717 (2004)

\bibitem{Rohe:1999sh}
D.~Rohe et~al., Phys. Rev. Lett. \textbf{83}, 4257 (1999)

\bibitem{Bermuth:2003qh}
J.~Bermuth et~al., Phys. Lett. \textbf{B564}, 199 (2003),
  \texttt{nucl-ex/0303015}

\bibitem{Schiavilla:2001qe}
R.~Schiavilla, I.~Sick, Phys. Rev. \textbf{C64}, 041002 (2001),
  \texttt{nucl-ex/0107004}

\bibitem{Lung:1992bu}
A.~Lung et~al., Phys. Rev. Lett. \textbf{70}, 718 (1993)

\bibitem{Kubon:2001rj}
G.~Kubon et~al., Phys. Lett. \textbf{B524}, 26 (2002), \texttt{nucl-ex/0107016}

\bibitem{Anklin:1994ae}
H.~Anklin et~al., Phys. Lett. \textbf{B336}, 313 (1994)

\bibitem{Markowitz:1993hx}
P.~Markowitz et~al., Phys. Rev. \textbf{C48}, 5 (1993)

\bibitem{Bruins:1995ns}
E.E.W. Bruins et~al., Phys. Rev. Lett. \textbf{75}, 21 (1995)

\bibitem{Anklin:1998ae}
H.~Anklin et~al., Phys. Lett. \textbf{B428}, 248 (1998)

\bibitem{Xu:2000xw}
W.~Xu et~al., Phys. Rev. Lett. \textbf{85}, 2900 (2000),
  \texttt{nucl-ex/0008003}

\bibitem{Gao:1994ud}
H.~Gao et~al., Phys. Rev. \textbf{C50}, 546 (1994)

\bibitem{Xu:2002xc}
W.~Xu et~al. (Jefferson Lab E95-001), Phys. Rev. \textbf{C67}, 012201 (2003),
  \texttt{nucl-ex/0208007}

\bibitem{Rock:1982gf}
S.~Rock et~al., Phys. Rev. Lett. \textbf{49}, 1139 (1982)

\bibitem{Glazier:2004ny}
D.I. Glazier et~al., Eur. Phys. J. \textbf{A24}, 101 (2005),
  \texttt{nucl-ex/0410026}

\bibitem{Herberg:1999ud}
C.~Herberg et~al., Eur. Phys. J. \textbf{A5}, 131 (1999)

\bibitem{Eden:1994ji}
T.~Eden et~al., Phys. Rev. \textbf{C50}, 1749 (1994)

\bibitem{Ostrick:1999xa}
M.~Ostrick et~al., Phys. Rev. Lett. \textbf{83}, 276 (1999)

\bibitem{Madey:2003av}
R.~Madey et~al. (E93-038), Phys. Rev. Lett. \textbf{91}, 122002 (2003),
  \texttt{nucl-ex/0308007}

\bibitem{Reichelt:2003iw}
T.~Reichelt et~al. (Jefferson Laboratory E93-038), Eur. Phys. J. \textbf{A18},
  181 (2003)

\bibitem{Plaster:2005cx}
B.~Plaster et~al. (Jefferson Laboratory E93-038), Phys. Rev. Lett. \textbf{91},
  122002 (2003), \texttt{nucl-ex/0511025}

\end{thebibliography}

\end{document}